\documentclass{aa}  
\usepackage{eqnarray}
\usepackage{caption,subcaption}

\newcommand{\molec}{H$_2$CO\,}

\usepackage{graphicx}
\usepackage{color}
\usepackage{txfonts}
\begin{document}

   \title{The SiO outflow from IRAS 17233-3606 at high resolution.}

   \author{P.D. Klaassen\thanks{pamela.klaassen@stfc.ac.uk}
          \inst{1,2} \and
          K.G. Johnston \inst{3,4} \and
          S. Leurini \inst{5} \and
          L.A.  Zapata \inst{6} 
          }

   \institute{Leiden Observatory, Leiden University, PO Box 9513, 2300 RA Leiden, The Netherlands
   \and UK Astronomy Technology Center, Royal Observatory Edinburgh, Blackford Hill, Edinburgh EH9 3HJ, UK
   \and Max Planck Institute for Astronomy, K\"onigstuhl 17, 69117, Heidelberg, Germany \and
   Department of Physics and Astronomy, University of Leeds, Leeds, LS2 9JT, UK
   \and    Max Planck Institut f\"ur Radioastronomie, Auf dem H\"ugel 69, 53121, Bonn, Germany   \and    Centro de Radioastronom\'ia y Astrof\'isica, UNAM, Apdo. Postal 3-72 (Xangari), 58089 Morelia, Michoac\'an, M\'exico     
       }

   \date{Received August 8, 2014; accepted December 18, 2014}

  \abstract
   {Jets and outflows are key ingredients in the formation of stars across the mass spectrum. In clustered regions, understanding powering sources and outflow components poses a significant problem.}
   { To understand the dynamics in the outflow(s) from a cluster in the process of forming massive stars.}
   {We use new VLA observations of the molecular gas (SiO, CS, OCS and \molec) in the massive star forming region IRAS 17233-3606 which contains a number of HII regions.  We compare these observations to previously published molecular data for this source in order to get a holistic view of the outflow dynamics.}
   {We find that the dynamics of the various species can be explained by a single large scale ($\sim 0.15$ pc) outflow when compared to the sizes of the HII regions, with the different morphologies of the blue and red outflow components explained with respect to the morphology of the surrounding envelope. We further find that the direction of the velocity gradients seen in OCS and \molec are suggestive of a combination of rotation and outflow motions in the warm gas surrounding the HII regions near the base of the large scale outflow.}
   {Our results show that the massive protostars forming within this region appear to be contributing to  a single outflow on large scales. This single large scale outflow is traced by a number of different species as the outflow interacts with its surroundings. On the small scales, there appear to be multiple mechanisms contributing to the dynamics which could be a combination of either a small scale outflow or rotation with the dynamics of the large scale outflow.}


   \maketitle
%

\section{Introduction} 

Although high mass (M$>$ 8 M$_\odot$) stars may form in isolation \citep{dewit04}, most form in clusters with an associated initial mass function (IMF) of lower mass stars.  Indeed, it is the clustered nature of high-mass star-formation that is often cited as one of the key hurdles (along with large distances and high radiation pressures) in our understanding of the formation of these stars \citep{ZY07}.

In attempts to overcome the hurdles of large distances, many studies focus on the large scale structures (i.e. outflows) and generally attribute the largest outflow to the most massive star in the region. Once this large scale outflow is identified, efforts are subsequently concentrated for finding signatures of a rotating disk. In this manner, many smaller scale rotating structures have been found surrounding massive (proto) stars \citep[i.e.][]{Beuther09,K09, Beltran11}

There has long been speculation about the nature of outflows from massive star forming regions, specifically as to why their energetics seem to scale  super-linearly with stellar mass from the low mass regime \citep[i.e.][]{DC13}.  Some argue that opening angles widen with age or stellar mass \citep[e.g.][]{Shepherd05}, while others suggest precession may lead to lowering the collimation \citep[e.g.][]{Raga09}, and there are still further theories  suggesting  how the competing outflows from the various accreting protostars interacting with each other produce the large scale wide outflows seen in high mass regions.  This is in addition to the mechanisms responsible for explosive events like those seen in Orion BN/KL \citep{Zapata09_orion, Zapata10}

IRAS 17233 is a nearby  \citep[1kpc,][]{Leurini13} high-mass star forming region containing multiple HII regions of various sizes and likely evolutionary states. In this study, we focus on the emission from, and dynamics of the molecular gas associated with the HII regions in the VLA1 and VLA2 regions as defined by \citet{Zapata08b}.

In VLA1, the spectral energy distributions (SEDs) of the continuum sources are consistent with optically thin HII regions \citep{Zapata08b}. In VLA2, the SEDs of the brightest objects (VLA2a,b) indicate Hypercompact (HC) HII regions, while the SEDs of the VLA2c,d could either be interpreted as due to HCHII regions or highly dusty disks.

Previous observations of the molecular gas in this region have concentrated primarily on the CO J=2-1 \citep[i.e][at 5.4$\times$1.9$''$ ]{Leurini09} and SiO J=5-4 \citep[e.g][at 3.2$\times$2.5$''$]{Leurini13} emission, and have identified three distinct outflows coming from the VLA1-2 region.  The multiple outflows were identified by being spatially distinct, and the emission coming from distinct velocity intervals \citep{Leurini08}.  

In this study, we incorporate these observations with new high resolution Karl G. Jansky Very Large Array (VLA) observations of SiO, CS, and other species to present a new interpretation of the outflow(s) from this system, as well as the smaller scale dynamics in the VLA1-2 region.

\begin{figure*}
\includegraphics[width=\textwidth]{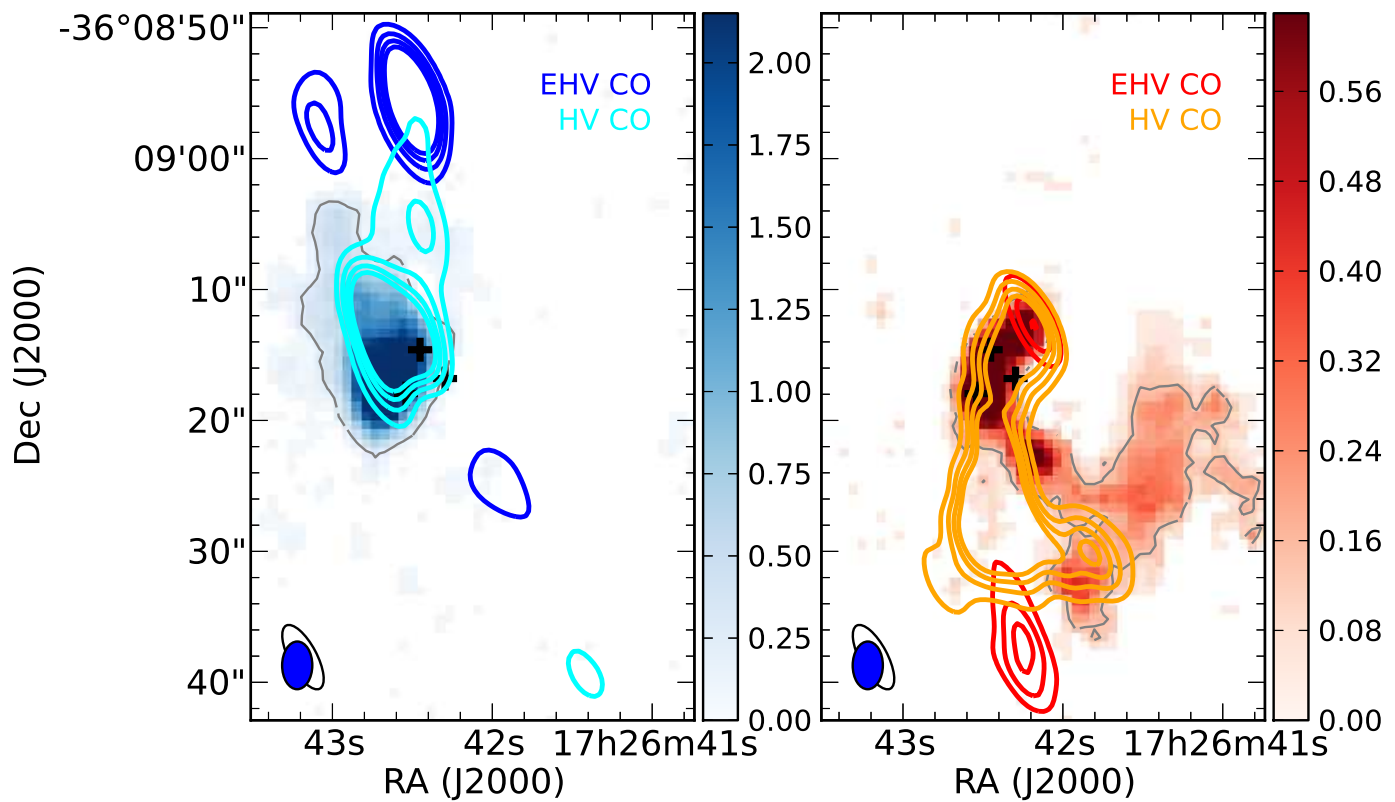}
\caption{CS integrated intensity with contours of CO emission.  {\it Left:} EHV blue shifted CO emission (blue contours) from \citet{Leurini09} starting at 3 $\sigma$ and increasing at intervals of 3$\sigma$. The HV velocity blue shifted CO is shown in cyan contours starting at 5$\sigma$ and increase at intervals of 5$\sigma$.  {\it Right} EHV red-shifted CO emission (red contours) from \citet{Leurini09} starting at 3 $\sigma$ and increasing at intervals of 3$\sigma$. The HV red shifted CO is shown in orange contours starting at 10$\sigma$ and increase at intervals of 10$\sigma$.  The colour scales show the blue and red shifted CS emission integrated over velocities greater than 4 km s$^{-1}$ (red or blue shifted) with respect to the LSR velocity, and has units of Jy beam$^{-1}$ km s$^{-1}$.  The thin grey contour shows the 3$\sigma$ levels of the CS integrated intensities (4.6$\times10^{-2}$ and 8.1$\times10^{-2}$ Jy beam$^{-1}$ km s$^{-1}$ for red and blue respectively). The synthesised beams for the CS (blue, filled) and CO (black, open) observations are given in the bottom left corner of both panels.The units of the scale bars are Jy beam$^{-1}$ km s$^{-1}$.}
\label{fig:CS_CO_outflow}
\end{figure*}

\section{Observations}  

Here we present VLA observations of IRAS 17233-3606 which was observed on 27 May, 2013 in the CnD hybrid configuration with the Q band receiver. There were eight narrowband spectral windows, and eight wideband. The narrowband windows had a spectral resolution of 125 kHz and bandwidth of 32 MHz (256 channels), while the wideband windows had a spectral resolution of 2 MHz and bandwidth of 128 MHz (64 channels). The total execution time of the observations was four hours, with 136 minutes on source. J1256-0547 was used as the bandpass calibrator and J1626-2951 was used as the gain calibrator. The flux calibrator was 3C286.

The data were calibrated in CASA \citep{CASA}, self calibrated on the continuum, and natural weighting was used when imaging.  The continuum emission was taken from the eight wideband spectral windows, and has an rms noise of 0.37 mJy/beam in a 2.80$''\times$ 1.69$''$ beam at a position angle of 29.5$^\circ$. SiO, CS, \molec, and OCS were observed for this source, and the transitions, observing frequencies, rms noise limits and beam sizes are listed in Table \ref{tab:molecules}. The rms noise limits in these spectral line cubes are all approximately 2 mJy/beam in each 0.78 km s$^{-1}$ channel.  

\begin{table}
\caption{Observed Transitions and beam properties}
\begin{tabular}{rcccc}
\hline \hline
Species & Transition & Freq. & {Beam} &Rms Noise\\
&& (GHz) &($''\times'',^\circ$) & (mJy/beam)\\ 
\hline
SiO & J=1-0 & 43.425&2.0$\times$1.6, -9& 1.8\\
OCS & J=4-3 & 48.655 &3.4$\times$1.8, 1&8.8\\
\molec & 4$_{1,3}$ - 4$_{1,4}$ &48.285 & 2.7$\times$1.8, -3&5.5\\
CS & J=1-0 & 48.995 & 3.6$\times$2.3,-0.3 & 12\\
\hline
\end{tabular}
\tablefoot{The rms noise limits are for a single (0.78 km s$^{-1}$) channel, and the quoted beam angle is the position angle of the major axis. }
\label{tab:molecules}
\end{table}

\section{Results}  

\subsection{Large scale outflow traced by CS and SiO}
\label{sec:results_sio}

CO, CS, and SiO are good tracers of outflow dynamics, but for very different reasons.  CO is good for tracing the high velocity molecular outflow because of its high abundance. Nearer the source velocity, the line becomes optically thick and we can only trace surface layer dynamics. CS, a less abundant molecule, is better for tracing the higher density gas when CO becomes optically thick. In the ISM, Si is frozen out onto dust grains, but can be liberated through the passage of shocks,  allowing it to be observed \citep[e.g][]{Gusdorf08}.  Thus, a combination of observations of these three species gives an excellent view of molecular outflows and how they interact with their environment.

Of the species directly observed for this study, CS is the most extended. The blue shifted emission to the North-East has an extent of $\sim12''$, while the red shifted emission to the South-West has an extent of $\sim25''$.  Our observations are sensitive to structures $\lesssim 60''$, and so it is unlikely  we are missing flux in these observations.  No single dish observations of CS J=1-0 exist for this source, so we cannot directly confirm this.  However, IRAS17233 was observed in CS J=2-1 by \citet{Bronfman96}.  Using the online RADEX calculator \citep{vandertak07}, and an ambient temperature of 30 K, we find good agreement (to within $\sim20\%$) between these two sets of observations and the expected line ratio at this temperature.

 The extended emission regions of CS correspond well to the `HV' CO ([-130,-25] and [16,50] km s$^{-1}$) emission regions presented in \citet{Leurini09}, and plotted with CS in Figure \ref{fig:CS_CO_outflow}. The velocities of the CS gas here are a factor of 2.5 lower than those of the `HV' gas suggesting these species are respectively tracing layers of higher and lower density within the same outflow.  It is interesting to note that the two blue `EHV' CO  ([-200,-130] and [90,120] km s$^{-1}$) peaks appear at the ends of the CS and `HV' CO outflow components.  These similarities between the CS and CO emission indicate that these two species are tracing the same outflow, only highlighting the gas at different densities and velocities. in Table \ref{tab:outflow} we present the outflow dynamics calculated from the CS emission presented here.  We note that while CS does appear to trace much of the outflowing material, it is only through the combination of observations of multiple tracers that we are able to piece together the full extent of the outflow. Thus, since the CS does not trace the full outflow, the derived outflow parameters are lower limits to the true mass, and energy of the outflow itself.  The outflow dynamics were calculated using the same methods described in \citet{Peters2014b} or \citet{Klaassen13a} using an ambient temperature of 30K, and a CS abundance of 10$^{-6}$ \citep{Shirley03}. The age of the outflow was estimated by dividing the extent of the CS emission by its mean velocity, and then used to calculate the outflow luminosity and mass loss rate.
 
\begin{figure}
\includegraphics[width=\columnwidth]{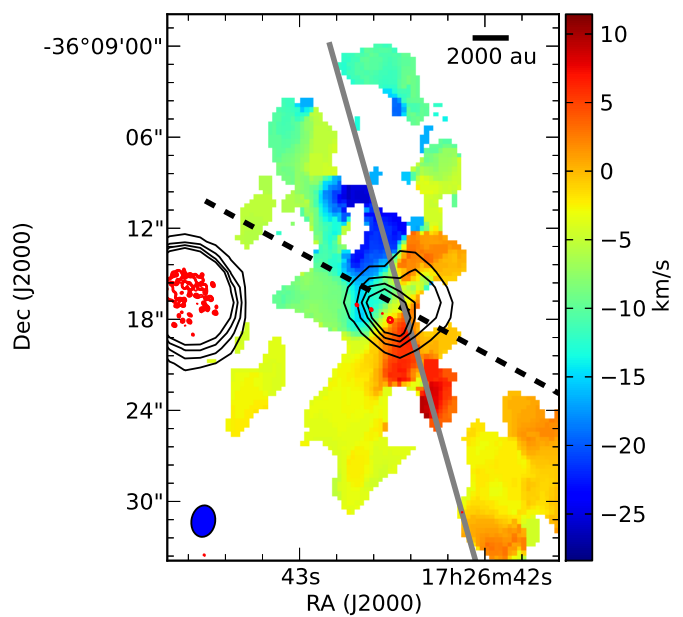}
\caption{Intensity weighted velocity (first moment) map of the SiO emission coming from the VLA1-2 region. The black contours show the 48 GHz continuum emission at 10, 20, 30 and 40$\sigma$ ($\sigma$ = 0.37 mJy/beam). The red contours show the 1.3 cm emission starting at 5 $\sigma$ and continuing in 1$\sigma$ intervals to 10$\sigma$. The red and black contours to the left show the VLA5 HII region. The solid grey and dashed black lines show the directions of the large scale outflow and small scale velocity gradients (respectively).  }
\label{fig:simple_sio}
\end{figure}

\begin{figure*}
\includegraphics[width=\textwidth]{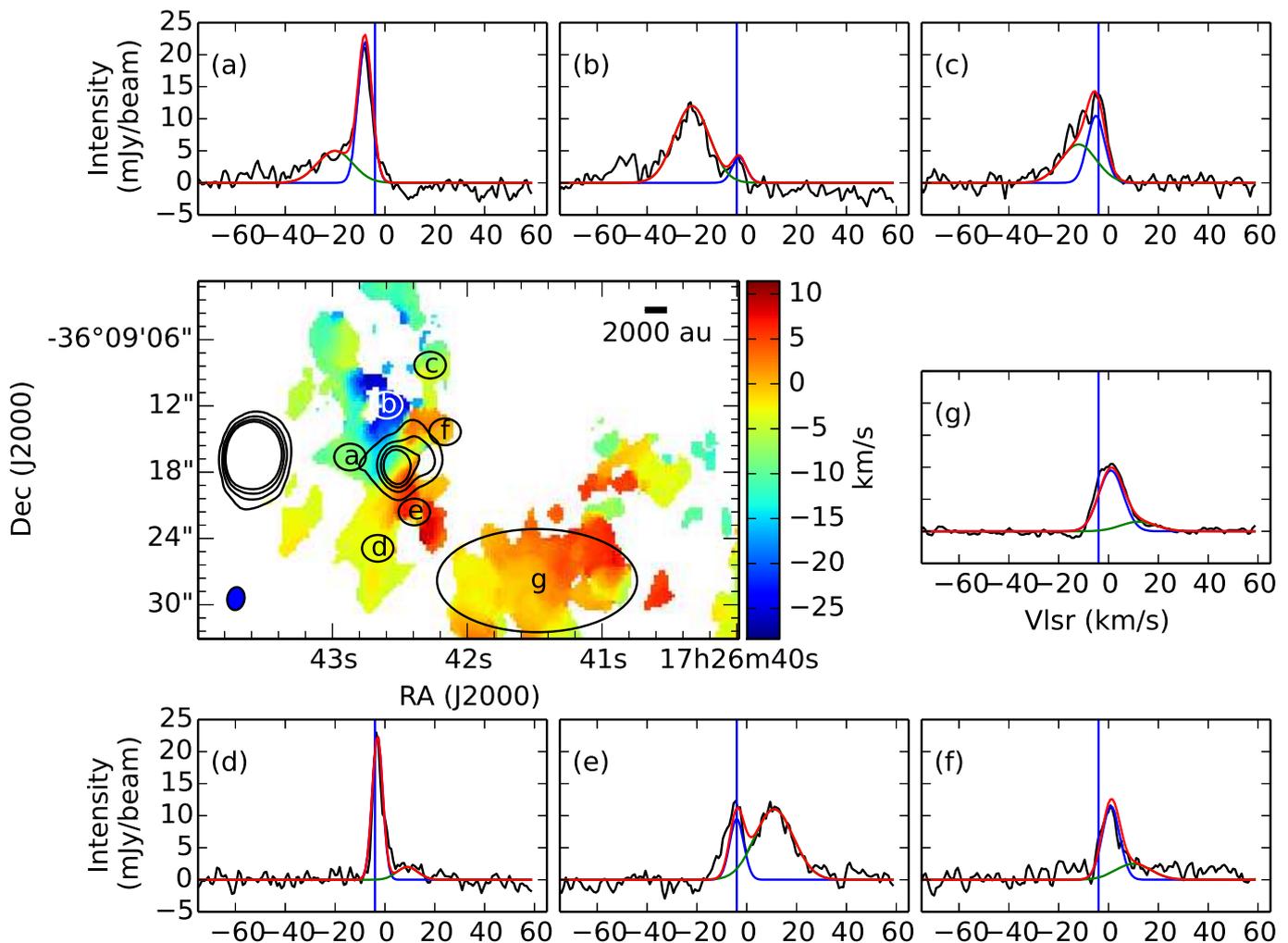}
\caption{The colour scale in the central panel shows the SiO intensity weighted velocity map as in Figure \ref{fig:simple_sio}.  The seven ellipses (labelled a-g) shown in this middle panel correspond to the regions over which the spectra in the correspondingly labelled panels were averaged.  The vertical blue line in each spectral panel corresponds to the V$_{lsr}$ of the source (-3.4 km s$^{-1}$). The blue and green curves in these panels correspond to the narrow, low velocity (blue) and broad, high velocity (green) components of the outflow, both red and blue shifted from the rest velocity.  }
\label{fig:sio_spectra}
\end{figure*}

The SiO however, appears to be tracing two distinct shock components.  The first moment map of SiO presented in Figure \ref{fig:simple_sio} shows highly blue and red shifted emission close to VLA1-2, with lower velocity gas wrapped around it.  Examining the spectral decomposition shown in Figure \ref{fig:sio_spectra} highlights these two components.  In both the red and blue shifted lobes, there exists a narrow (FWHM $\sim5$ km s$^{-1}$), low velocity (3-5 km s$^{-1}$ from V$_{\rm LSR}$) component as well as a broad (FWHM $\sim10$ km s$^{-1}$), high velocity (16 km s$^{-1}$ from V$_{\rm LSR}$) component.  When analysing these gaussian components, the peak velocities and gaussian widths were allowed to vary slightly (they were held to within 1 km s$^{-1}$), while the amplitudes allowed to freely vary.  
 
\begin{figure}
\includegraphics[width=\columnwidth]{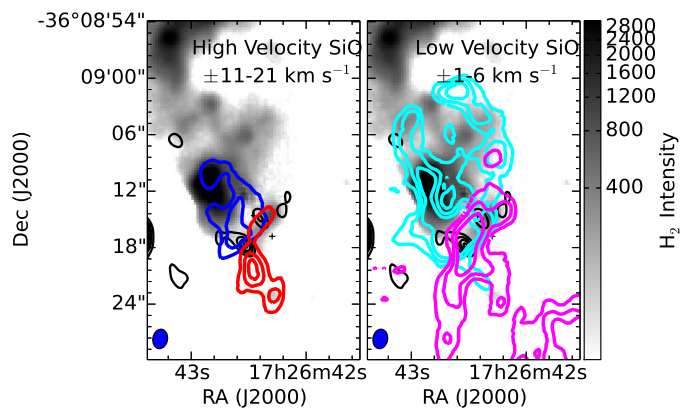}
\caption{Blue- and red-shifted integrated intensities of the high (left) and low (right) velocity SiO emission components outlined in Figure \ref{fig:sio_spectra}. The SiO emission is superimposed on the H$_2$ emission from the VLA1-2 region (greyscale).  In each case, the contours of the SiO emission are set at 10, 20, 30 and 40 $\sigma$.  From these images, the spatial distribution of the two components can be seen, with the lower velocity shocks tracing the edges of the outflow cavity, and the high velocity shocks tracing the powered jet/outflow close to the HII regions in the centre of the emission.}
\label{fig:sio_components}
\end{figure}

Using the Gaussian peak velocities and their widths, we created the integrated intensity maps presented in Figure \ref{fig:sio_components}. This suggests that  there are two sets of shocks in this region; high velocity near the outflow core, and lower velocity where the outflow is interacting with the envelope.

\begin{table*}
\begin{center}
\caption{Lower Limits to Outflow Properties derived from CS emission}
 \begin{tabular}{rl@{$\pm$}rl@{$\pm$}rl@{$\pm$}rl@{$\pm$}ll@{$\pm$}ll@{$\pm$}l}
\hline
\hline
&\multicolumn{2}{c}{Mass} & \multicolumn{2}{c}{Momentum}& \multicolumn{2}{c}{Energy} & \multicolumn{2}{c}{Luminosity} & \multicolumn{2}{c}{Mass Loss Rate}\\
&\multicolumn{2}{c}{($M_\odot$)} & \multicolumn{2}{c}{($M_\odot$ km s$^{-1}$)}& \multicolumn{2}{c}{(10$^{45}$ erg)} & \multicolumn{2}{c}{($L_\odot$)} & \multicolumn{2}{c}{($10^{-3}M_\odot$ yr$^{-1}$)}\\
\hline
 Blue& 30.104 & 0.003  & 305.53 & 0.05 & 35.70 & 0.01 & 36 & 2 & 3.7 & 0.2\\
 Rd& 12.347 & 0.003  & 233.79 & 0.08 & 45.00 & 0.02 & 46 & 3 & 1.5 & 0.1\\
 Total& 42.451 & 0.004  & 539.32& 0.09 & 80.70 & 0.02 & 82 & 4 & 5.3 & 0.2\\
\hline
\end{tabular}
\label{tab:outflow}

\end{center}
\end{table*}

\begin{table*}
\begin{center}
\caption{Gaussian fits to the peaks of the CS, OCS and \molec emission, and resultant velocity gradients and dynamical masses.}
\begin{tabular}{rrrr@{$\times$}lrrrrr}
\hline \hline \\
Species & \multicolumn{2}{c}{Position of peak} &\multicolumn{2}{c}{Semi Axes ($''\times''$)} & &\multicolumn{2}{c}{Fluxes} & Velocity Gradient & Mass\\
& RA & DEC & Major & Minor   & P.A. ($^\circ$)  & Integrated & Peak & (km s$^{-1}$ pc$^{-1}$) & M$_\odot$\\ \hline
CS & 17:26:42.6& -36:09:16.86&3.7 & 3.3 & 163 & 61.90 & 8.98 & - & -\\
OCS &17:26:42.5 & -36:09:17.47& 3.8&3.2 & 101& 16.98 & 5.33 & 189$\pm$12 & 275\\
\molec & 17:26:42.5 &-36:09:17.10& 3.5 & 3.4 &22 & 12.05 & 3.45 & 232$\pm$13 & 250\\
\hline
\end{tabular}
\tablefoot{The uncertainties of the axis lengths and position angles are 0.03$''$ and 3$^\circ$ respectively, and come from the deconvolved gaussian fits to the data. The integrated and peak fluxes have units of Jy beam$^{-1}$  km s$^{-1}$ and Jy beam$^{-1}$, respectively.}
\label{tab:gaussians}
\end{center}
\end{table*}

\subsection{Small scale dynamics}
\label{sec:small_scale}

\begin{figure}
\includegraphics[width=\columnwidth]{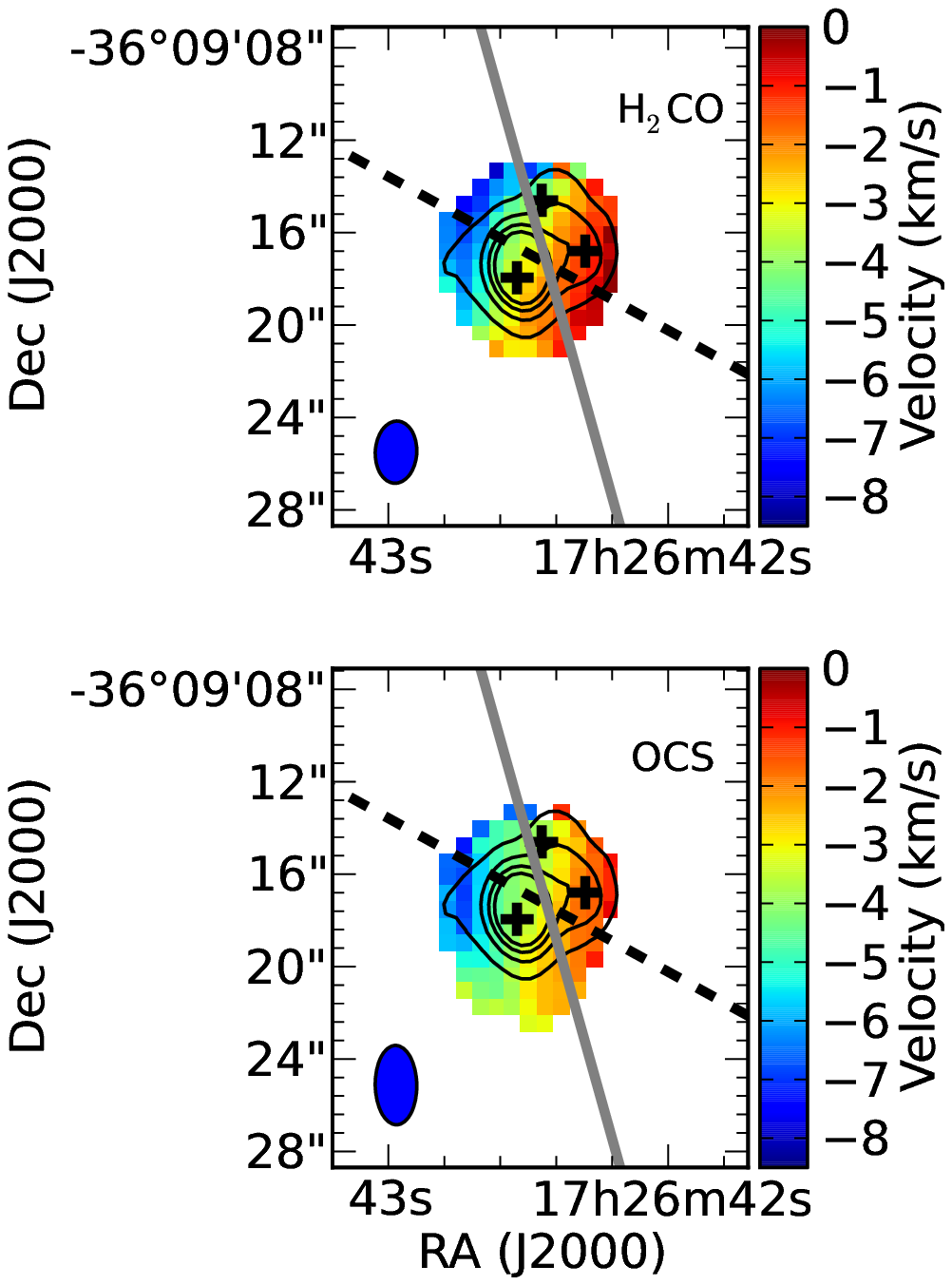}
\caption{Intensity weighted velocity (first moment) maps of the OCS and \molec emission observed towards VLA 2a, VLA1a and VLA1b. The contours show the 10, 20, 30, and 40 $\sigma$ contours of the 48 GHz continuum emission. The solid grey line shows the position angle of the outflow as outlined in Figure \ref{fig:simple_sio}, and the dashed black line shows the direction of the velocity gradients indicated here. The colour scales in each panel are identical, which indicates that the \molec has a larger velocity gradient than the OCS, since they emit on roughly the same spatial scales.}
\label{fig:moment_one}
\end{figure}

Tracing the large scale outflow to its powering source(s) is impossible with the these data because other motions appear to dominate the small scale ($\lesssim0.01$pc) emission.  The direction of the velocity gradient on small scales is different than on the large scales, as shown with lines in Figures \ref{fig:simple_sio} and \ref{fig:moment_one}.

Both CS and SiO are strongly peaked in the central region concentrated on VAL1-2. Indeed, to its 50\% emission contour, CS appears broadly gaussian, the parameters of which are given in Table \ref{tab:gaussians}.

On these smaller scales, all of the molecules observed for this study have velocity gradients along the same direction, tilted by 45$^\circ$ from the position angle of the outflow (see the dashed and solid lines in Figures \ref{fig:simple_sio} and \ref{fig:moment_one}).  We will focus our analysis of these velocity gradients on OCS and \molec which only sit on these smaller scales. The conditions required to excite these species are denser and hotter than the other observed tracers making them ideal for understanding these dynamics.

OCS is generally only found when temperatures rise above 200 K \citep{Jiminez-serra12}. \molec is formed on dust grains, and thus requires warm temperatures to be liberated.  Their abundances can also be quite low in comparison to CS and CO. The abundance of OCS is observed to be $\sim2.2\times10^{-8}$ with respect to H$_2$ \citep{Charnley97}, and the abundance of \molec varies between 10$^{-9} - 10^{-10}$ between regions \citep[e.g.][]{Ginard12,Noble12,Guzman11}. The combination of high excitation temperatures and low abundances means that both OCS and \molec trace the inner (warm) areas near the HII region.

Both OCS and \molec emit in roughly Gaussian regions, the parameters of which are given in Table \ref{tab:gaussians}.  From these fits, we see that this emission is well resolved by our observations. The velocity gradients calculated from the velocity shifts in the emission ($\sim 8$ km s$^{-1}$) and the sized of the emitting regions are listed in Table \ref{tab:gaussians}, as are the internal masses which would be required to support these velocity gradients if they were due to purely Keplerian rotation.

That these velocity gradients are at 45$^\circ$ to the outflow and not perpendicular as would be expected from a traditional disk/outflow system suggests that there is some other mechanism contributing to these dynamics.  The nature of this emission is discussed further in Section \ref{sec:desc-rot}.

\begin{figure}
\includegraphics[width=\columnwidth]{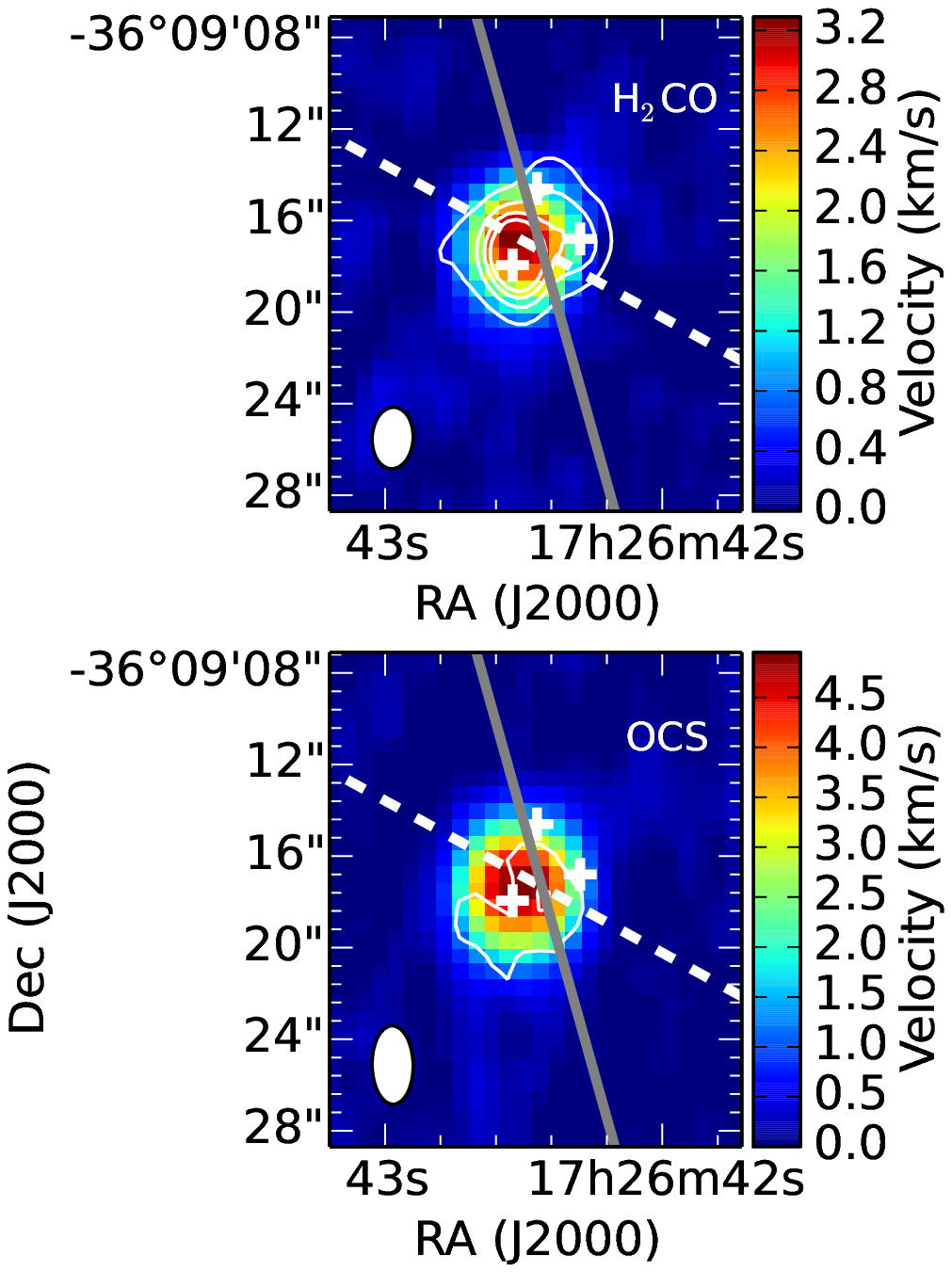}
\caption{Integrated intensity  (zeroth moment) maps of the OCS and \molec emission observed towards VLA 2a, VLA1a and VLA1b (white crosses). In the \molec plot, the contours show the 10, 20, 30, and 40 $\sigma$ contours of the 48 GHz continuum emission, while in the OCS plot, they show the 3 and 6 $\sigma$ contours of the O$^{13}$CS emission. The solid grey line shows the position angle of the outflow as outlined in Figure \ref{fig:simple_sio}, and the dashed white line shows the direction of the velocity gradients indicated here.}
\label{fig:moment_zero}
\end{figure}

Using the methods described in \citet{K09}, we calculated the gas mass from the emission of OCS.  To do this, required a number of assumptions. The first, that the OCS is optically thin, comes from comparing the OCS intensity to that of O$^{13}$CS where the O$^{13}$CS peaks in our observations.  The second is that the temperature of the OCS gas is 300 K, stemming from requiring a minimum temperature near 200 K for the emission to appear. Varying the temperature by 100 K up or down only changes the results at the level of 10\%.  The third assumption is that the abundance of OCS is 2.2$\times10^{-8}$ \citep{Charnley97}.  From these assumptions, we derived a gas column density, which we multiplied by the emitting area delimited by the semi-major and -minor axes listed in Table \ref{tab:gaussians}.  Through this analysis, we determine a gas mass of $\sim$690 M$_\odot$ in the OCS emitting region.

\section{Discussion}

\subsection{Large scale outflow}

\begin{figure}
\includegraphics[width=\columnwidth]{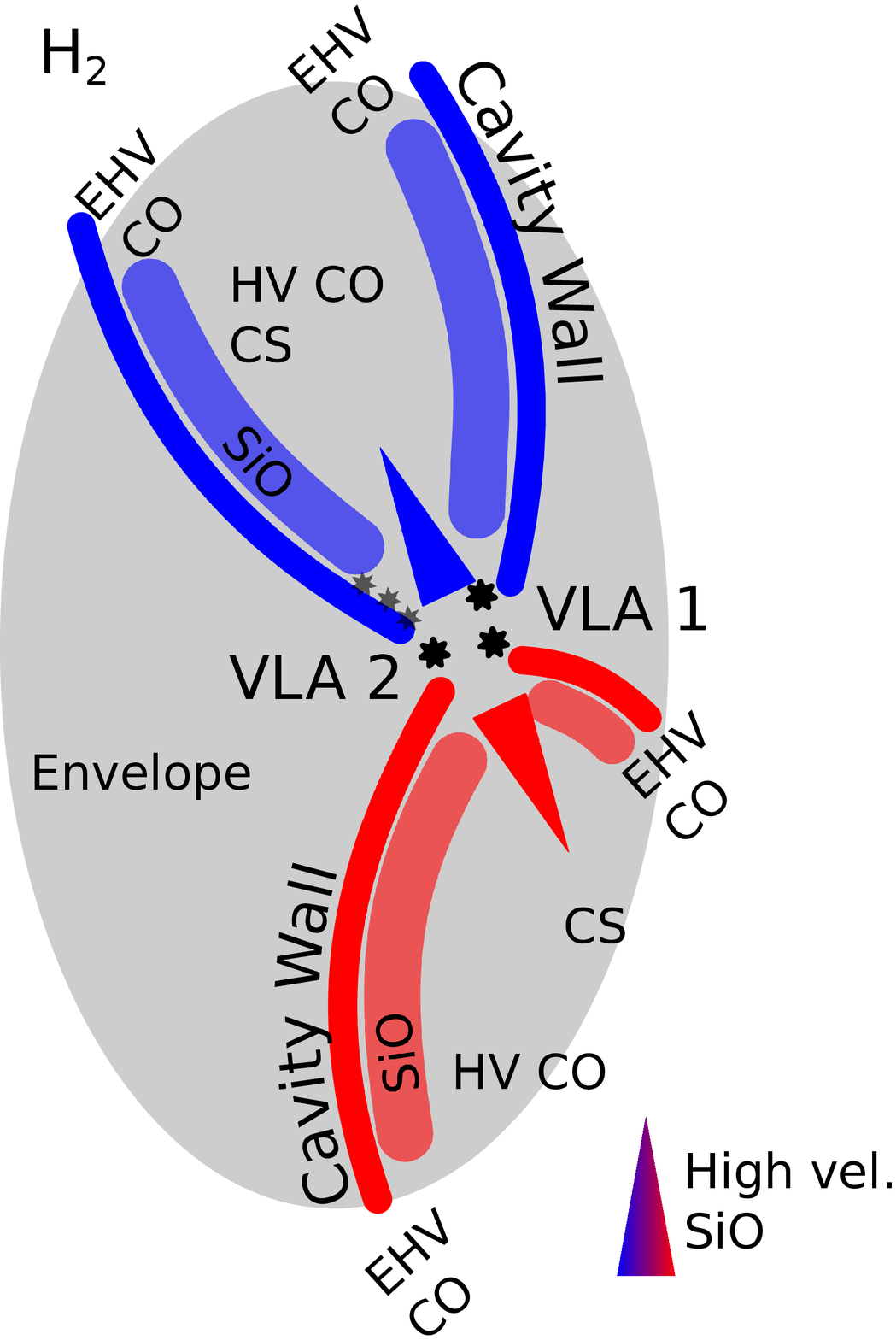}
\caption{Cartoon of the IRAS17233 outflow, as it appears on the sky. The black stars in the centre of the outflow show the approximate positions of VLA2a, VLA1a and VLA1b, while the smaller (grey) stars show the positions of VLA 2b-d.  The cones at the centre of the figure show the high velocity SiO component, while the wide lines within the cavity wall show the low velocity SiO emission component.  The regions of EHV and HV CO, as well as general CS emission are labelled, and the envelope through which the outflow is expanding is shown with a grey oval.  Note that the ends of the outflow extend further than the envelope, explaining the presence of EHV CO at the ends of the outflowing structure. The presence of an H$_2$ emission feature to the North of the outflow is also depicted here.}
\label{fig:cartoon}
\end{figure}

Contrary to previously published high resolution observations of this region \citep[e.g.][]{Leurini08,Leurini11,Leurini13}, we suggest that there is only one large scale outflow from the VLA1-2 region of IRAS 17233, and that the various molecules detected in this region are simply tracing its various components.  A simple cartoon of the various outflow components is given in Figure \ref{fig:cartoon} to help guide the reader through the following arguments.

Here we first discuss how the various species are tracing a single outflow in the blue-shifted lobe, and then extend our analysis  to the more complicated red-shifted lobe.

In this picture, the highest velocity SiO emission is tracing the shocks produced, centrally, by the infall/accretion driven outflow.  The lower velocity SiO is in turn tracing the shocks produced as the outflow hits the relatively high density material at the edges of the outflow cavity.

\begin{figure}
\includegraphics[width=\columnwidth]{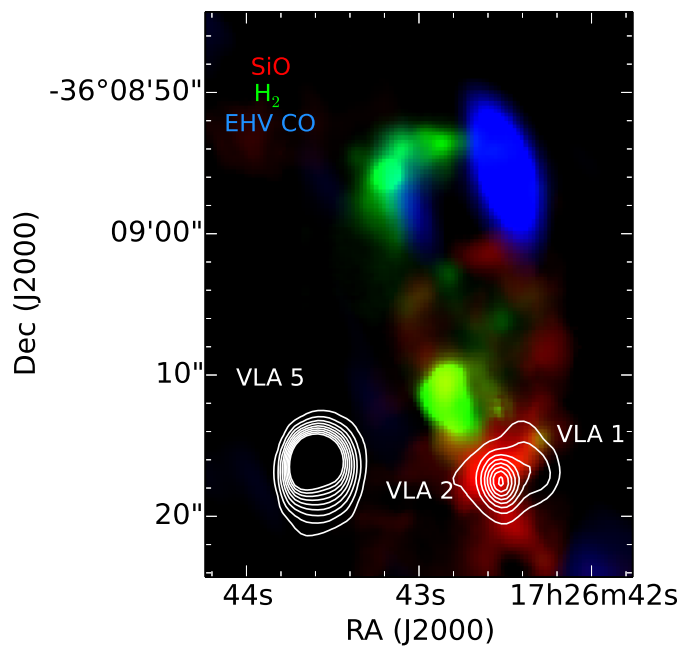}
\caption{ SiO (red), CO (blue) and H$_2$ (green) emission in the blue shifted outflow lobe from IRAS17233 VLA1-2.  The SiO and CO emission are integrated over [-40,0] and [-200,-130] km s$^{-1}$ respectively. The H$_2$ is integrated over the entire line. This shows how the different emission components all contribute to a single outflow. The contours show the 48 GHz continuum emission, with VLA5 to the left, and the VLA1-2 region to the right.}
\label{fig:blue_lobe}
\end{figure}

The `HV' CO component as well as the CS emission are both tracing the bulk of the outflow emission within the well defined outflow cavity (delineated by the lower velocity SiO emission) as shown in Figure \ref{fig:CS_CO_outflow}.  At larger distances from the powering sources in VLA1-2, where the CS and `HV' CO emission starts dropping off, the 2.12 $\mu$m H$_2$ emission shown in Figures \ref{fig:sio_components} and \ref{fig:blue_lobe} begins to become stronger.

This detection of H$_2$ as the CS and CO emission becomes weaker suggests lower extinctions, and by extension, lower density molecular gas. This simultaneously explains why the CO and CS emission fades further from the source: at the edges of the envelope from which the stars in VLA1-2 are forming.

As the outflow continues outward, it encounters more rarefied gas. As the outflow moves though the lower density material outside of the shocked envelope, it begins impacting onto this material and entraining it. This explains the two `EHV' CO emission features at the tips of the low velocity SiO shocks. The conditions in the `EHV' region must not be right for the outflow to create C-shocks (and form SiO), but it is still present and entraining CO.

Further still from VLA1-2 is a bright clump of H$_2$ emission (towards the top of Figures \ref{fig:sio_components} and \ref{fig:blue_lobe}, in greyscale and green, respectively). This could be tracing the area where the outflow is hitting the (relatively) static ambient medium. Indeed, the H$_2$ at the head of the outflow has a bowed morphology.

Thus, the blue shifted emission of various molecules to the North-East of VLA1-2 can thus be explained by a single outflow interacting with an environment with changing characteristics.

Interpreting the redshifted emission is not as straight forward. However, using the assumption that there is a large scale outflow in this system, the various components of the red-shifted emission can be explained.

The `HV' CO to CS relationship is not quite as clear in the red-shifted outflow lobe.  Near the central core of VLA1-2, and along the (narrow) core of the outflow, the two species appear to be tracing the same gas, however the bulk of the `HV' CO emission tends to bend southward from the main outflow direction once it has left the central core.

 Simultaneously, there is `EHV' CO emission even further South than the `HV' CO.  Again, as with the blue-shifts outflow lobe, there is a clear edge to the CS emission where the `EHV' CO emission begins (See Figure \ref{fig:CS_CO_outflow}).  Here, we suggest that this CO emission (at higher velocities than the CS) may be tracing the less dense gas surrounding the VLA1-2 region; that the outflow has been able to punch through the edge of the envelope at this location. This then leads to the conclusion that the CS is tracing entrained envelope material, and the `HV' CO is tracing the outflow material that has not been slowed down by the envelope.  This conclusion also explains the red `EHV' CO near the core of the VLA1-2 system. The red-shifted outflow has managed to break free of the envelope (there is no CS emission), and expand freely, at high velocities, into the low density ambient medium.  This `early' breakout into the ambient medium of the red-shifted outflow lobe explains why there is no double walled outflow cone in the red-shifted material that mirrors the one seen in the low velocity blue-shifted SiO.

\begin{figure}
\includegraphics[width=\columnwidth]{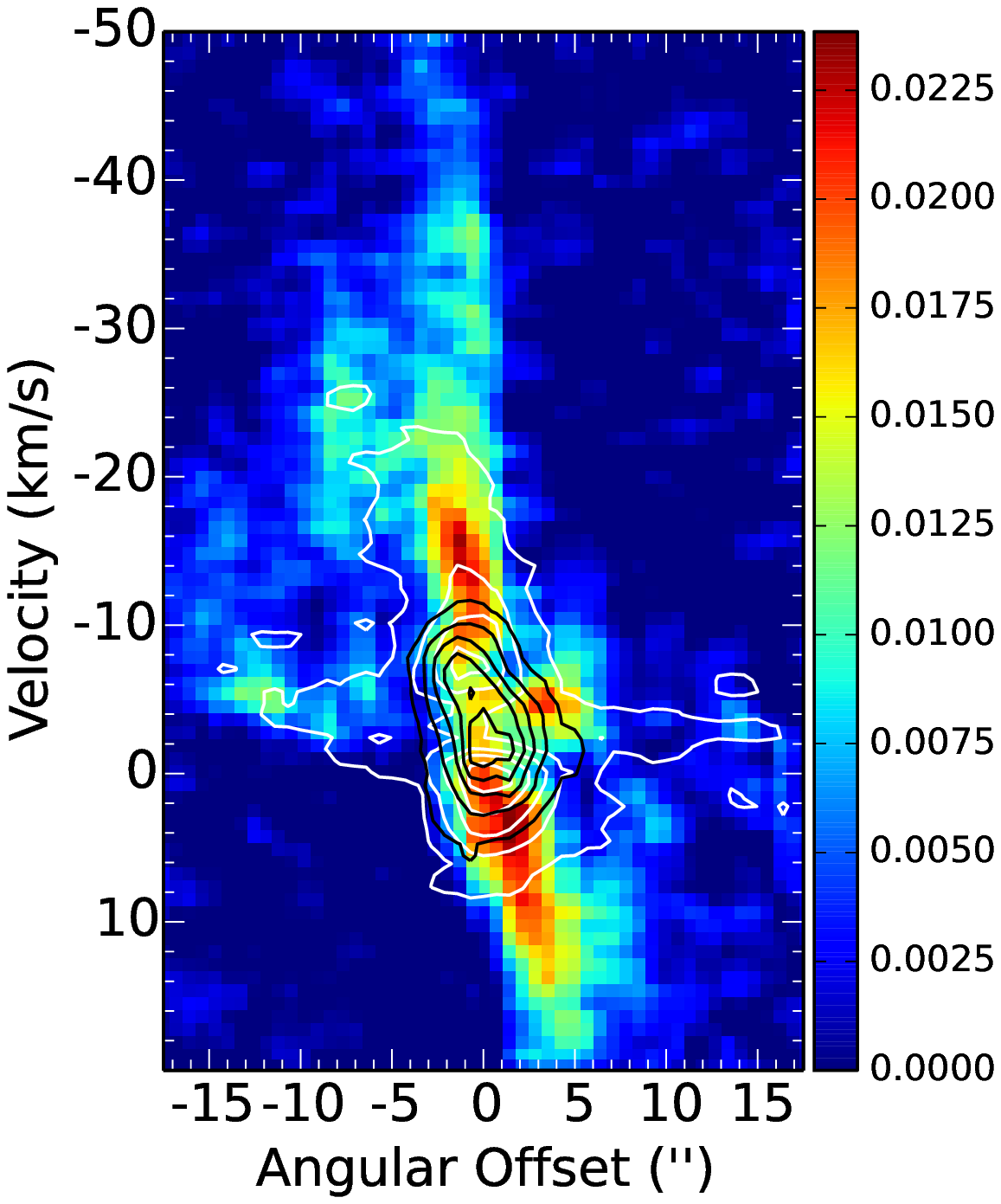}
\caption{ SiO (colourscale), CS (white contours) and \molec (black contours) position-velocity diagram for a cut along the grey line shown in Figures \ref{fig:simple_sio} and \ref{fig:moment_one}. The units of the colour scale are Jy/beam, and the contours of the CS and \molec emission correspond to 10\% to 90\% of the peak value, in steps of 20\%. The SiO emission in particular shows both high and low velocity components, while the CS traces an arc with both high and low velocity gas. These patterns are best seen in the blue shifted material.}
\label{fig:CS_SiO_PV} 
\end{figure}

In Figure \ref{fig:CS_SiO_PV} we show the Position Velocity diagram of the CS and SiO emission along the direction of the solid grey line in Figures \ref{fig:simple_sio} and \ref{fig:moment_one}.  We see a butterfly morphology in the SiO emission reminiscent of both the jet and outflow structures seen in CO and SiO (respectively) in the low mass region L1448 by \citet{Hirano10}. Our SiO shows both of these components, in a single tracer.  In the SiO emission (colourscale) we see both the low and high velocity components of the SiO emission.   Because SiO is a good tracer of shocked gas \citep[see, for instance,][]{Gusdorf08}, these two SiO components are likely tracing two different shock regions associated with this outflow. The first, high velocity shock region, is showing the energetic outflow coming from the protostars, while the second, low velocity shock region, is showing the interaction of the outflow with the cavity wall.  That these various components are symmetric about the VLA1-2 region in terms of red and blue shifted material, suggests we are observing a single outflow and its interaction with its environment.
 
What is causing this single, large scale, outflow is not completely clear.  If there is a single massive forming star responsible for this outflow in either VLA1 or VLA2, we are unable to distinguish it using our observations.  The VLA1-2 region appears to be embedded in a rotating structure spanning a few thousand au, and the powering source for the large scale outflow is embedded within it.

Alternatively, \citet{Peters2014b} suggest a scenario in which multiple massive protostars may be collectively contributing to the one large scale outflow observable in massive star forming regions.  In their scenario, the protostars form within a rotating envelope, and form in a relatively co-planar manner.  They find that there is a single `collective' outflow produced in their simulations, and that the contributions from the different protostars cannot be disentangled.  It is possible that we are observing one of these `collective' outflows in IRAS 17233, however we cannot exclude the simpler explanation of a single and dominant protostar being responsible for the single large scale outflow.

 \subsection{Small scale dynamics: a combination of rotation and outflow}
 \label{sec:desc-rot}
 
While we cannot rule out that there is a small scale outflow at a 45$^\circ$ to the large scale one \citep[OF1 in the nomenclature of ][]{Leurini08}, here we explore the possibility of the small scale velocity gradient resulting from a combination of rotation and outflow motions.

 As shown in Figures \ref{fig:moment_one} and \ref{fig:moment_zero}, the emitting regions for OCS and \molec are spatially similar, and both are quite circular despite being well resolved with our VLA beam.  The emission peaks are separated by 0.3$''$ (300 AU), which is significant given our beam size ($\sim 2"$) and signal to noise ratio ($\sim 500$) for these species.  We note that both peaks lie along the line joining VLA2a and VLA1b, as well as the axis of the velocity gradient.  The velocity gradients of the emission from these two species have the same position angle ($\sim 60^\circ$, see Figure \ref{fig:moment_one}) and are therefore probably tracing similar gas components.  The position angle (highlighted by a dashed black line in Figures \ref{fig:simple_sio} and \ref{fig:moment_one}) is at about a 45$^\circ$ angle to the large scale outflow direction (highlighted with a solid grey line in the same figures).  That there is a 45$^\circ$ angle, instead of the small scale velocity gradients being parallel or perpendicular to the large scale outflow suggests that OCS and \molec are tracing a combination of velocity structures.
 
 \citet{Leurini11} suggest that the observed velocity gradient seen in their CH$_3$CN (see their Figure 9) was likely due to the combined effects of two outflows.  \citet{Beuther09}, who observed this velocity gradient in ammonia, attributed it to (non-Keplerian) rotation of the envelope from which the cluster is forming.
 
 We propose a common ground between these two explanations. We suggest that these species, as well as the CH$_3$CN and OCS observed by \citet{Leurini11}, are tracing a combination of outflow and rotational motions.  This proposal is based on comparing the observed velocity gradients (including their position angles) to toy radiative transfer models similar to those presented in \citet{KSWang12}.
 
 \citet{KSWang12} showed that when both rotation and outflow motions are included in radiative transfer modelling, the position angle of the velocity gradient shifts depending on the relative velocities of these two components.  We applied similar modelling techniques, as described below.
 
 Without tweaking outflow and rotation parameters to explicitly model IRAS 17233, we set up LIME \citep{LIME} radiative transfer models of our observed species. A brief discussion of the models can be found in Appendix \ref{sec:lime_models}
 
We find that to reproduce the position angle of the \molec and OCS velocity gradients requires rotational velocities approximately three times greater than outflow velocities at a characteristic radius of 1000 AU (1$''$).  On these size scales, we cannot separate out purely rotational or outflow motions, however we can take the characteristic SiO velocities discussed in Section \ref{sec:results_sio} as a proxy, and calculate rotational dynamics from that.  The characteristic velocities of the SiO (5 and 16 km s$^{-1}$ with respect to the rest velocity of the source) correspond to rotational velocities of 15 and 48 km s$^{-1}$, which we further assume hold at 1000 AU.  If the gas were undergoing Keplerian motions, we would expect internal masses of 250 and 2600 M$_\odot$ for the two SiO components. The first is a third of the gas mass derived above from the OCS emission, and the second is larger than the gas mass of the entire region.  If this velocity were achieved at radii differing from 1000 AU, the mass would increase (for larger radii) or decrease (for smaller radii) linearly with the radius. To reach the 690 M$_\odot$ calculated for the gas mass would require a characteristic radius of 3000 AU, which is only slightly smaller than the radius of the OCS emission.
 
 \section{Conclusions}

 We have presented molecular line observations of the gas in the massive star forming region IRAS 17233-3606, and find that  despite there being multiple HII regions in the VLA1 and VLA2 regions \citep[as defined by][]{Zapata08b}, there is a single large scale ($\sim$ 0.15 pc) outflow coming from these HII regions. These outflow motions are traced by SiO, CS and previously published CO and H$_2$ emission in this region.  Our results show that despite there being multiple sites of massive star formation within VLA1-2, there is likely only one large scale outflow from the system. The large scale SiO emission appears to break down into two components; a small scale highly collimated, high velocity emission component, and a larger scale, lower velocity component.  The first component traces shocks produced by the powered outflow itself, while the second, larger scale component traces the shocks at the edges of the outflow cavity.
 
 The emission from OCS and \molec are on much smaller scales, and the velocity gradients seen in their emission could be due to a combination of a rotating envelope around VLA1-2, and material being swept up into the large scale outflow traced by SiO, CS and CO.  We find that the implied masses from the small scale velocity gradients ($\sim 250$ M$_\odot$) are consistent with forming a few massive stars, and are are approximately a third of   the mass determined from the integrated intensity of the molecular emission ($\sim 690$ M$_\odot$).

\begin{acknowledgements}
The authors would like to thank J.C. Mottram for helpful discussions during the preparation of this manuscript, and both the referee and the editor for their comments which helped strengthen our results and conclusions. This manuscript makes use of VLA data. The National Radio Astronomy Observatory is a facility of the National Science Foundation operated under cooperative agreement by Associated Universities, Inc.
\end{acknowledgements}

\bibliographystyle{aa}

\appendix

\section{Description of Toy Radiative Transfer Models}
\label{sec:lime_models}

Without attempting to directly model the velocity field of IRAS17233, we used LIME radiative transfer models to determine what combination of outflow and rotational motions could cause a 45$\circ$ degree offset between the large scale outflow and the small scale velocity gradients in the molecular gas.

\begin{figure}
\begin{centering}
\begin{subfigure}{0.4\textwidth}
\includegraphics[width=\textwidth]{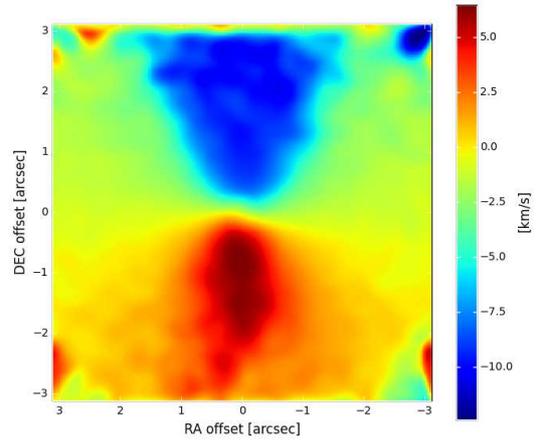}
\caption{outflow motions}
\end{subfigure}
\begin{subfigure}{0.4\textwidth}
\includegraphics[width=\textwidth]{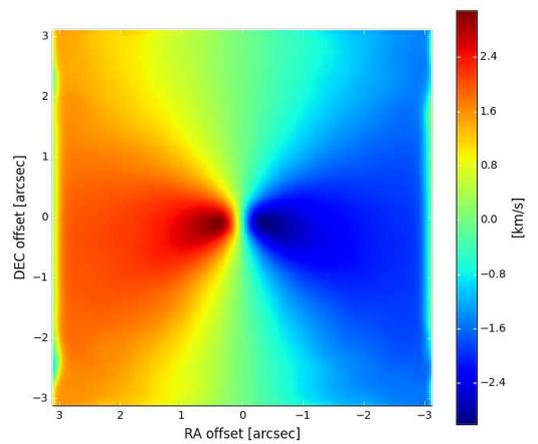}
\caption{rotational motions}
\end{subfigure}
\begin{subfigure}{0.4\textwidth}
\includegraphics[width=\textwidth]{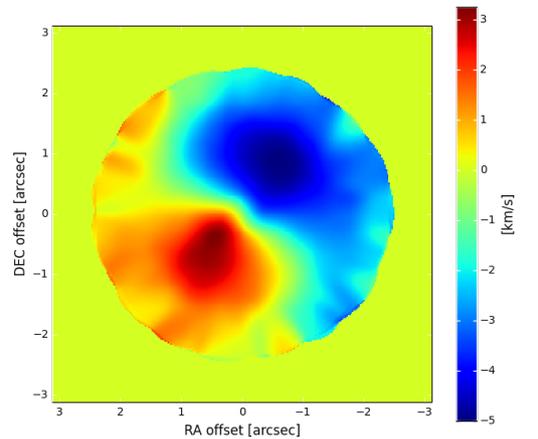}
\caption{combined motions}
\end{subfigure}
\caption{Toy radiative transfer models showing how rotational and outflow motions can combine to produce a velocity gradient that is halfway between the directions of the two dominant dynamical processes.  From top to bottom, the panels show outflow motions only, rotational motions (perpendicular to the outflow) and the combined effects of the two when the outflow velocities are one third those of the rotational velocities at a radius of 1000 au.}
\label{fig:lime_models}
\end{centering}
\end{figure}

To do this, we simplistically assumed temperature, density and velocity profiles of the form:

\begin{eqnarray*}
T = T_o\left(\frac{r}{r_d}\right)^{-\beta} & V_r = v_{r_o}\left(\frac{r}{r_o}\right) \\
n = n_o\left(\frac{r}{r_d}\right)^{-\alpha}(\sin{\theta})^f + n_o*\left(\frac{r}{r_d}\right)  &
V_\phi = \sqrt{\frac{GM}{r_{20au}}}
\end{eqnarray*}

where  $r$, $\theta$ and $\phi$ represent the coordinate grid, $\beta=1$, $\alpha=1.5$, $f$ is the flattening parameter of the rotating structure ($f=5$), $r_d$ is the distance between VLA2a and VLA1 (2000 au), and  $r_{20au}$ is the given radius outwards of 20 au, but fixed at 20 au for smaller radii to avoid a divide by zero.  The resultant velocity grids from this modelling are shown in Figure \ref{fig:lime_models}. As shown here, a combination of purely rotational motion along the midplane, and outflow motion perpendicular to it creates a scenario where the observed combination of these two effects is a velocity gradient at 45$^\circ$ to both. The outer edges of the combined motions shows where the density and temperature become low enough that OCS and \molec are no longer observable.

\section{SiO and CS Channel maps}

\begin{figure*}
\begin{center}
\includegraphics[width=0.75\textwidth]{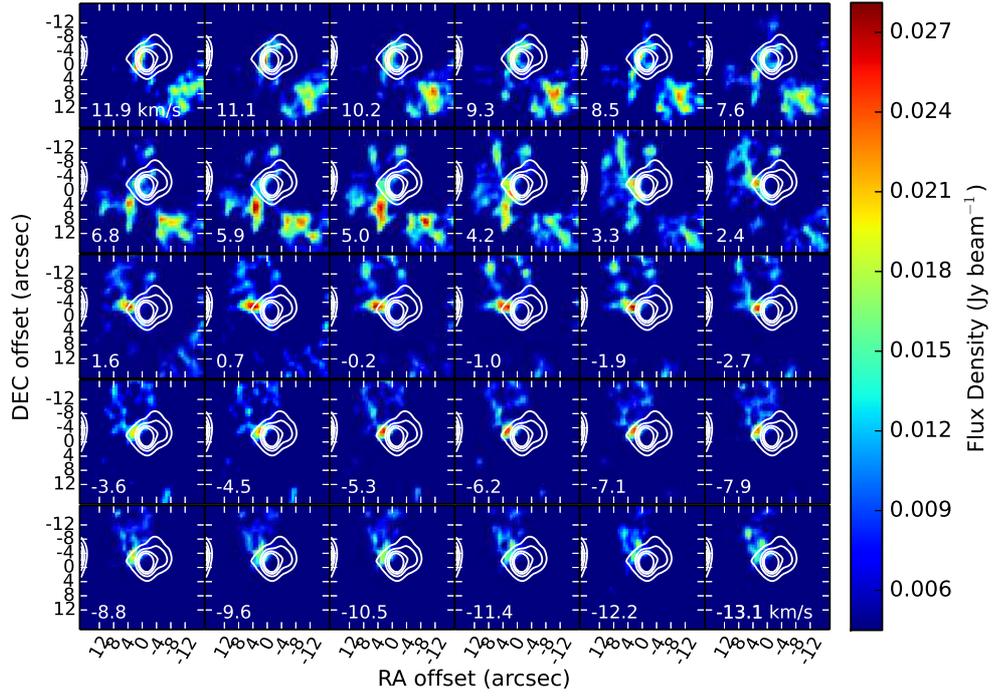}
\caption{ SiO channel map. The white contours indicate the 48 GHz continuum, consistent with previous figures, and the colourscale shows the SiO emission starting at 3$\sigma$}
\label{fig:sio_chan}
\end{center}
\end{figure*}

\begin{figure*}
\begin{center}
\includegraphics[width=0.75\textwidth]{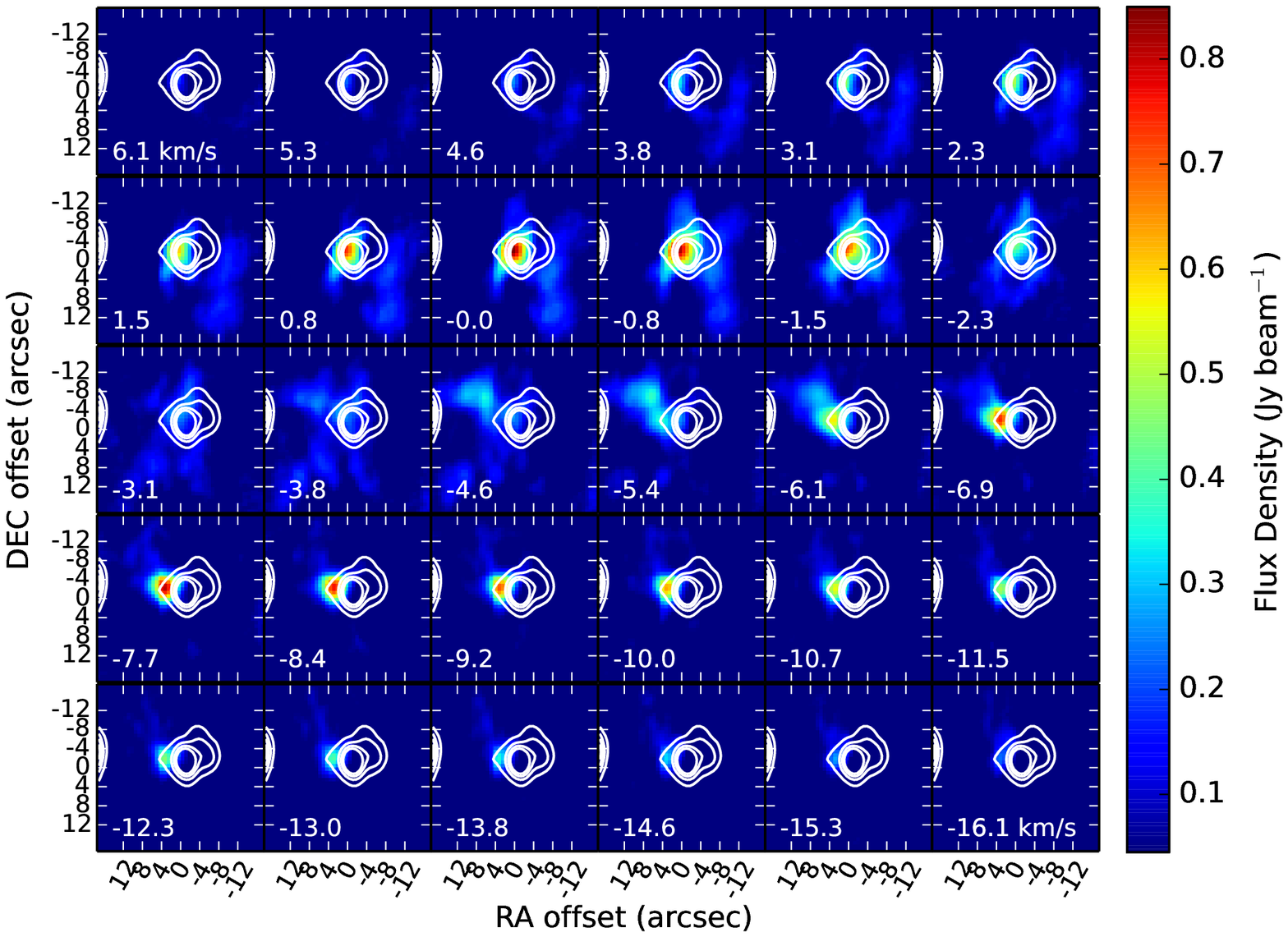}
\caption{ CS channel map. The white contours indicate the 48 GHz continuum, consistent with previous figures, and the colourscale shows the CS emission starting at 3$\sigma$}
\label{fig:cs_chan}
\end{center}
\end{figure*}

Figures \ref{fig:sio_chan} and \ref{fig:cs_chan} show, respectively, the channel maps of the SiO and CS emission surrounding IRAS17233 VLA2-1. The centre of each panel shows the same 48 GHz continuum contours as previous figures for reference.  Note that the sizes of each panel are consistent between the two figures, however the starting and ending velocities are slightly shifted to reflect the bulk of the emission for each of the two species. 

\end{document}